\documentstyle[sprocl,epsfig]{article}
\input{psfig}
\def\be{\begin{equation}}
\def\ee{\end{equation}}
\def\bea{\begin{eqnarray}}
\def\eea{\end{eqnarray}}

\begin{document}
\begin{flushleft}
\vspace*{-3cm}
Interner Bericht
\\
DESY-Zeuthen 96--01
\\
Januar 1996
\vspace*{3cm}
\end{flushleft}

\title{
SEMI-ANALYTICAL APPROACH TO OFF-SHELL $W$ AND $Z$ \\
  PAIR PRODUCTION
\footnote{Contribution to the Proceedings of the Workshop on Physics and
Experiments with Linear Colliders, Sept 8--12, 1995, Morioka-Appi,
Iwate, Japan}  
}
\author{ 
D. BARDIN, D. LEHNER, \underline{T. RIEMANN}
}
 
\address{
Deutsches Elektronen-Synchrotron DESY
\\
Institut f\"ur Hochenergiephysik IfH, Zeuthen
\\
Platanenallee 6, D-15738 Zeuthen, Germany
}
 
\author{ A. LEIKE}
 
\address{
Ludwig-Maximilians-Universit\"at M\"unchen, Sektion Physik
\\
Theresienstra{\ss}e 37, D-80333 M\"unchen, Germany
}
 
 
\maketitle\abstracts{
We give a short review of the results for four fermion production,
which have been obtained by the semi-analytical approach.
The angular degrees of freedom (typically five or more) are integrated
over analytically while the integrations over invariant fermion pair
masses (typically two or more) remain to be performed by numerical
methods.
In addition to doubly resonating cross sections from virtual two boson
production, QED corrections and background contributions were
determined.
However, a large variety of final state topologies has not been
treated so far.   
}
 
\section{Introduction
\label{intro}}
Among the most interesting processes to be studied at a high energy
linear collider is pair production of gauge and Higgs bosons.
Since these heavy particles are instable, one has to study
experimentally their decay products, namely four final state
fermions:
\bea
e^+ e^- \rightarrow (W^+W^-, ZZ, Z\gamma, \gamma \gamma, ZH, \ldots)
\rightarrow f_1 {\bar f}_2 f_3 {\bar f}_4 .
\label{1}
\eea
After integration over five angular variables, the total cross section
for reactions~(\ref{1}) may be generically written as follows:
\bea
  \sigma^{res}(s) & = &
  \int ds_1 \rho_B(s_1)
  \int ds_2 \rho_B(s_2)
  \;\; \sigma_{0}(s;s_1,s_2)
  \label{sigww}
\eea
where $s_1 = (p_1+p_2)^2$ and $s_2= (p_3+p_4)^2$.
The bosons' Breit-Wigner densities
\begin{eqnarray}
  \rho_B(s_{i}) & = & \frac{1}{\pi} \frac {M_B \Gamma_B}
   {|s_{i} - M_B^2 + i M_B \, \Gamma_B |^2}
  \times {\mathrm {BR}}
   \label{rhow}
\end{eqnarray}
attain the following narrow width limit:
\begin{equation}
  \rho_B(s_{i}) \stackrel {\Gamma_B \rightarrow 0} {\longrightarrow}
  \delta(s_{i} - M_B) \times {\mathrm {BR}}.
  \label{normal}
\end{equation}

The expressions for $\sigma_{0}(s;s_1,s_2)$ have been derived for off-shell
$W^+W^-$ production in~\cite{muta},
for $ZZ, Z\gamma, \gamma \gamma$ production in~\cite{teupitz},
for $ZH$ production in~\cite{gentle_nc24h} (see also references
therein).
In section~\ref{CC11} we will give explicit examples for the basic
cross section $\sigma_0(s;s_1,s_2)$.

Naturally, the four fermion final states in~(\ref{1}) are produced not
only by doubly resonant amplitudes, but also by many singly resonant
and non-resonant tree level background amplitudes, which are
characterized by different intermediate states.
In addition, radiative corrections must be accounted for.
Here, we will concentrate on results which have been obtained
by use of the semi-analytical method: 
QED initial state radiative corrections and background contributions. 
\subsection{Background\label{bg}}
The doubly resonant diagrams yield the dominant cross section
contributions even far above threshold; nevertheless, only together
with background gauge invariance may be achieved.  
The number of Feynman diagrams for a given process depends on the
topology of the final state. 
A classification has been introduced in~\cite{teupitz}:
\begin{itemize}
\item[CC]
Processes with final states of type $f_1^u{\bar f}_1^d f_3^u {\bar
  f}_3^d$ are called {\em charged current processes}:
{\tt CC11}, {\tt CC10}, {\tt CC09}; {\tt CC20}, {\tt CC18}.
\item[NC]
Processes with final states of type $f_1{\bar f}_1 f_3 {\bar f}_3$ are
called {\em neutral current processes}:
{\tt NC32}; {\tt NC24}, {\tt NC4$\cdot$16}, {\tt NC4$\cdot$12}, {\tt
NC4$\cdot$03}, {\tt NC10}, {\tt NC06}; {\tt NC48}, 
{\tt NC20}, {\tt NC21}, {\tt NC19}, {\tt NC12}, {\tt NC4$\cdot$36}, {\tt
NC4$\cdot$09}. 
\item[mix]
Processes which may be considered as both CC and NC types are
called {\em mixed processes}:
{\tt mix43}, {\tt mix19}; {\tt mix56}. 
An example for a mixed process is the production of
$u{\bar d} d{\bar u} \equiv u{\bar u} d{\bar d}$.
\end{itemize}
In addition, there are Feynman diagrams with Higgs bosons in the NC
and mixed cases.
The simplest background processes are those of the {\tt CC20} and the
{\tt NC24} classes. 
These have been studied in detail in~\cite{gentle_unicc11}
and~\cite{gentle_nc24,gentle_nc24h}, respectively.  
As an example, we discuss the {\tt CC11} process in
section~\ref{CC11}.

Tree level cross sections including backgrounds may be generically
written as
\be
  \label{totsig}
  \sigma^{\rm{Born}}(s) \; = \; \int d s_1 \int d s_2 \;\;
    \frac{\sqrt{\lambda}}{\pi s^2} \cdot
    \sum_k \frac{d^2 \sigma_k(s,s_1,s_2)}{d s_1 d s_2}
\ee
with $\lambda \equiv \lambda(s,s_1,s_2),~\lambda(a,b,c) =
a^2\!+\!b^2\!+\!c^2\!-\!2ab\!-\!2ac\!-\!2bc$ and 
\be
  \label{diffsig}
  \frac{d^2 \sigma_k}{d s_1 d s_2} \; = \;
    {\cal C}_k(s,s_1,s_2) \cdot {\cal G}_k(s,s_1,s_2)~~,
\ee
where ${\cal C}_k$ represents coupling constants and off-shell boson
propagators, while ${\cal G}_k$\ is a kinematical function obtained
after analytical angular integration.
The index $k$ labels cross section contributions with different
coupling structure {\em and} different Feynman topology.
\subsection{QED initial state radiation
\label{isr}}
The by far largest radiative corrections are due to QED initial state
radiation (ISR).
They may be described by the generic formula
\begin{eqnarray}
  \label{ISRxstot}
  \sigma^{ISR}(s) & = &
  \int d s_1 \int d s_2 \;\;
  \int d s' \;\;
    \sum_k \frac{d^3\Sigma_k(s,s';s_1,s_2)}{d s_1 d s_2 d s'}
\end{eqnarray}
with $s'=(p_1+p_2+p_3+p_4)^2$ and
\bea
  \label{ISRxsdif}
  \frac{d^3\Sigma_k(s,s';s_1,s_2)}{d s_1 d s_2 d s'} & = &
  ~~~{\cal C}_k \cdot
  \left[ \beta_e v^{\beta_e - 1} {\cal S}_k+{\cal H}_k \right]~~,
\eea
where $\beta_e = 2\,\frac{\alpha}{\pi} \, [ \ln (s/m_e^2) - 1]$\ ~and
$v = (1- s'/s)$.

The soft+virtual and hard corrections ${\cal S}_k$ and ${\cal H}_k$
separate into a universal, factorizing, process-independent and a
non-universal, non-factorizing, process-dependent part.
They are given by~\cite{gentle_nunicc,gentle_nuninc}
\bea
  {\cal S}_k(s,s';s_1,s_2) & = &
  \left[1 + {\bar S}(s) \right] \sigma_{k,0}(s';s_1,s_2)
  + \:\sigma_{{\hat S},k}(s';s_1,s_2) ~~~,
  \nonumber \\
  \hspace{-1cm}
  {\cal H}_k(s,s';s_1,s_2) & = &
  \underbrace{{\bar H}(s,s') \: \sigma_{k,0}(s';s_1,s_2)\;\;\;\;}_{Universal
    ~Part}
  + \underbrace{\sigma_{{\hat H},k}(s,s';s_1,s_2)}_{Non-universal~Part}
\eea
with $\sigma_{k,0}(s';s_1,s_2) \equiv \left[ \sqrt{\lambda}/(\pi s^{'2})
\right] \cdot 
{\cal G}_k(s',s_1,s_2)
$\
and the universal ${\cal O}(\alpha)$\ soft+ virtual and hard radiators
${\bar S}$\ and ${\bar H}$
\be
  {\bar S}(s) \; = \; \frac{\alpha}{\pi}
                    \left[  \frac{\pi^2}{3} - \frac{1}{2} \right]
                    + \frac{3}{4} \, \beta_e \hspace{1.7cm}
  {\bar H}(s,s') \; = \; - \frac{1}{2}
                         \left(1+\frac{s'}{s}\right)\beta_e~~.
\ee
If the index $k$ is associated with s-channel $e^+e^-$ annihilation
diagrams only,  only universal ISR contributions are present, because
non-universal ISR originates from the angular dependence of initial
state {\em t}- and {\em u}-channel propagators.
Non-universal ISR contributions are suppressed with respect to
universal ones, because they do not contain the leading logarithm
$\beta_e$.
They are, however, analytically very complex.
\section{Example 1: The {\tt NC8} process with
  complete initial state corrections
\label{NC08}}
\begin{figure}[t]
  \vspace*{1.6cm}
  \begin{center}
    \mbox{
      \hspace*{-2cm}
      \epsfysize=10.8cm
      \epsffile{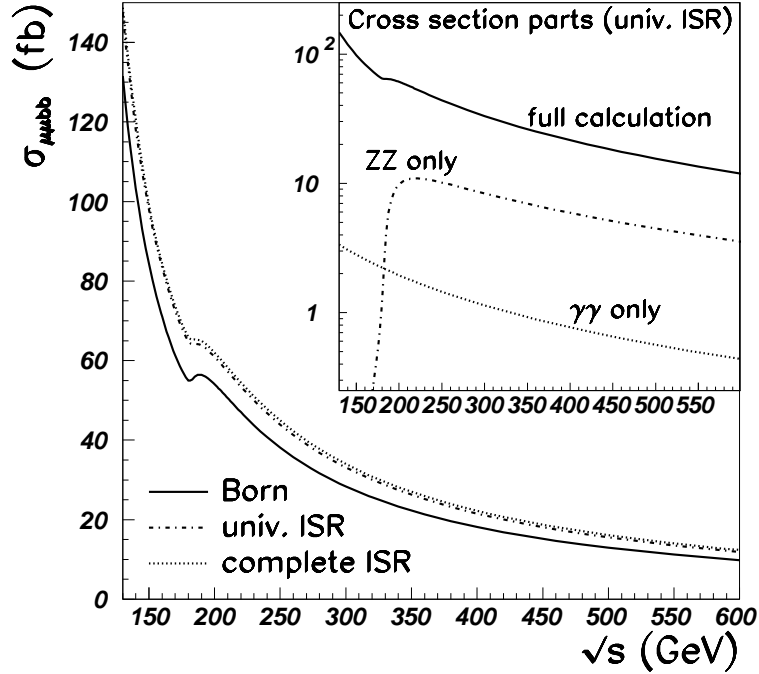}}
  \end{center}
  \vspace{-3.8cm}
  \caption[The neutral current pair cross section with ISR corrections]
    {\it The cross section for the {\tt NC8} process with ISR.}
  \label{figNC8}
\end{figure}

The {\tt NC8} process $ e^+e^- \rightarrow (ZZ,Z\gamma,\gamma\gamma)
\rightarrow f_1\bar{f_1}f_2\bar{f_2}
~~(f_1\!\neq\!f_2~,~f_i\!\neq\!e^\pm, \stackrel{_{(-)}}{\nu_e})$
is described by only one kinematical
function~\cite{teupitz,gentle_nc24},
\begin{eqnarray}
  {\cal G}_{\tt NC8}(s;s_2,s_2) & = &
s_1 s_2 \left[
    \frac{\,s^2+(s_1+s_2)^2\,}{s-s_1-s_2}\,{\cal L}(s;s_2,s_2) -
    2\right]~~, 
  \label{ZZmuta} \\
  {\cal L}(s;s_1,s_2) & = & \frac{1}{\sqrt{\lambda}} \,
  \ln\frac{s-s_1-s_2+\sqrt{\lambda}}{s-s_1-s_2-\sqrt{\lambda}}~.
\end{eqnarray}
Numerical results are presented in
figure~\ref{figNC8}~\cite{gentle_nuninc,gentle.f}.
%
%
As is seen from the figure, ISR corrections increase the cross section
considerably.
Universal contributions are of ${\cal O}$(+20\%), non-universal
contributions of ${\cal O}$(+3\%).
From the inset of the figure one concludes that most important cross
section contributions involve Z bosons {\em and} photons.
We mention that the doubly resonant Z pair cross section is easily
isolated by cuts on $s_1$\ and $s_2$.
\section{Example 2: The {\tt CC11} processes with
  initial state corrections
\label{CC11}}
The basic charged current four fermion process is described by three
kinematical functions~\cite{muta}:
\begin{eqnarray}
 \sigma^{\tt CC3}_0(s;s_1,s_2) & = &
 \frac{\left(G_{\mu} M_W^2 \right)^2} {8 \pi s} 
 \Biggl[  {\cal C}^s_{\tt CC3} \, {\cal G}^{33}_{\tt CC3}
        + {\cal C}^{st}_{\tt CC3} \, {\cal G}^{3f}_{\tt CC3}
        + {\cal C}^t_{\tt CC3} \, {\cal G}^{ff}_{\tt CC3} \Biggr].
  \label{sigww0}
\end{eqnarray}
The functions ${\cal G}_{\tt CC3}$ depend only on the virtualities:
\begin{eqnarray}
  {\cal G}^{ff}_{\tt CC3}(s;s_1,s_2) & = & \frac{1}{48}
    \left[ \rule[-.1cm]{0cm}{.5cm} \: \lambda +
           12 \, s \, \left( s_1 + s_2 \right) - 48 s_1 s_2 \right.
    \nonumber \\ 
& & \left. \hspace{1cm} \rule[-.1cm]{0cm}{.5cm}
           + 24 \left(s - s_1 - s_2\right) s_1 s_2
           \!\cdot\! {\cal L}(s;s_1,s_2)\right]~,
    \label{cc3gt} \\ \nonumber \\
  {\cal G}^{33}_{\tt CC3}(s,s_1,s_2) & = & \frac{\lambda}{192}
    \left[ \rule[-.1cm]{0cm}{.5cm} \lambda + 12\left(s s_1 + s s_2 +
           s_1 s_2 \right)\right]~,
    \label{cc3gs} \\ \nonumber \\
  {\cal G}^{3f}_{\tt CC3}(s;s_1,s_2) & = & \frac{1}{48}
    \left\{ \rule[-.2cm]{0cm}{.7cm} (s-s_1-s_2)
            \left[ \rule[-.1cm]{0cm}{.5cm}
                   \lambda + 12s(s s_1 + s_1s_2 +s_2 s) \right]
                 \right.
    \nonumber \\ & &  \left. \rule[-.2cm]{0cm}{.7cm} \hspace{.75cm}
      - 24 \left( s s_1 + s s_2 + s_1 s_2 \right) s_1 s_2
                   \!\cdot\! {\cal L}(s;s_1,s_2) \right\},
    \label{cc3gst}
\end{eqnarray}
Adding terms from background diagrams yields~\cite{gentle_unicc11}
\begin{equation}
  \label{CC11xsec}
  \sigma^{\tt CC11}(s) \; = \; \int d s_1 \int d s_2 \;\;
    \frac{\sqrt{\lambda}}{\pi s^2} \cdot
    \sum_{k=1}^{15} 
    {\cal C}_k \cdot {\cal G}_k(s,s_1,s_2)~.
\end{equation}
Using the symmetries of the process, the 15 functions may be expressed
by the three functions ${\cal G}_{\tt CC3}$, two of the process
{\tt NC24} -- ${\cal G}_{\tt NC8}, {\cal G}_{\tt
  NC24}$~\cite{gentle_nc24} -- plus only one new kinematical function,
which is the most complicated one:
\begin{eqnarray}
  {\cal G}_{\tt CC11}^{u,d} (s;s_1;s_2) \; = \;
     -120 s^4 \frac{s_1^3 s_2^3}{\lambda^3}
     {\cal L}(s_2;s,s_1) {\cal L}(s_1;s_2,s) \hspace{3.5cm}
     \nonumber
\end{eqnarray}
\vspace{-.6cm}
\begin{eqnarray}
 -s \Biggl[1+\frac{s(s\!-\!\sigma)}{\lambda}
     +20s^2\frac{s_1s_2}{\lambda^2}
     -30s^3s_1s_2\frac{s\!-\!3\sigma}{\lambda^3}\Biggr] \!
     \Biggl[s_1^2 {\cal L}(s_2;s,s_1)+s_2^2 {\cal L}(s_1;s_2,s)\Biggr]
     \; \nonumber \\
 -s(s_1\!-\!s_2) \Biggl[\frac{s\!-\!\sigma}{\lambda}
     +10s\frac{s_1s_2}{\lambda^2}
     -30s^2s_1s_2\frac{s\!+\!\sigma}{\lambda^3}\Biggr]
     \Biggl[s_1^2 {\cal L}(s_2;s,s_1)-s_2^2 {\cal L}(s_1;s_2,s)\Biggr]
     \;\; \nonumber \\
 -\frac{1}{12} \Biggl\{(s^2-\sigma^2)
     \Biggl[1+12\frac{s\sigma}{\lambda}-60s^2\frac{s_1s_2}{\lambda^2}\Biggr]
     \hspace{5.96cm} \nonumber
\end{eqnarray}
\vspace{-.6cm}
\begin{eqnarray}
 \hspace*{2cm}
 -8s_1s_2\Biggl[1-\frac{s(4s+5\sigma)}{\lambda}
     +15s^2\frac{s_1s_2}{\lambda^2}\Biggl(1-6\frac{s(s-\sigma)}{\lambda}
     \Biggr)\Biggr]\Biggr\},
\end{eqnarray}
with $\sigma \! \equiv \! s_1+s_2$. 

Numerical results have been produced with
programs~\cite{gentle_unicc11,wphact,wwgenpv,wto}.
Table~1 contains a comparison of the Born cross sections with
background in different channels for a wide range of beam energies.
At energies above the $ZZ$ threshold, the background modifies
$\sigma_{tot}$, but for all channels quite similarly. 
While the changes of the rates are of the percent level, this may be quite
different for other processes, e.g. those of the {\tt CC20} type.
\begin{table}[bt]
\caption[]
{\it
$\sigma_{tot}$ in pbarn for Born 4$f$ production as function of
$\sqrt{s}$ (in GeV). For this comparison, the 
branching ratios are {\em not} taken into account in the
single mode channels.  
\label{tcsia}
}
\begin{center}
\begin{tabular}{|c|r@{.}l|r@{.}l|r@{.}l|r@{.}l|}
\hline 
$\sqrt{s}$  
          &\multicolumn{2}{c|}{\tt CC3        }  
          &\multicolumn{2}{c|}{$L \nu l \nu$ } 
          &\multicolumn{2}{c|}{$l \nu qq'$   }
          &\multicolumn{2}{c|}{$QQ'qq'$      }
\\ \hline 
\hline
30   
&  \multicolumn{2}{c|}{1.4519$\times 10^{-7}$}  &  5&9295$\times
10^{-6}$   &  7&2478$\times 
10^{-6}$   &   5&0897$\times 10^{-6}$   
\\ \hline 
60   
&  \multicolumn{2}{c|}{1.9358$\times 10^{-5}$} &  4&0025$\times 10^{-5}$ &  4&6879$\times
10^{-5}$ &  3&5441$\times 10^{-5}$ 
\\ \hline 
91.189   &  0&11225  &  0&021329  &  0&024551  &  0&018975 
\\ \hline 
176  &16&225 &16&242 &16&243 &16&243
\\ \hline 
200  &18&578 &18&586 &18&588 &18&588
\\ \hline 
500  &        7&3731    &7&3301 &7&3318 &7&3334 
\\ 
{\tt WWGENPV}&7&3731(7) & \multicolumn{2}{c|}{}     &
\multicolumn{2}{c|}{}     &7&3332(5) 
\\ \hline 
1000         &2&9888   &2&9342 &  2&9344&2&9348 
\\ 
{\tt WWGENPV}&2&9887(7) & \multicolumn{2}{c|}{}     &
\multicolumn{2}{c|}{}     &2&9342(5) 
\\ \hline 
2000         &1&5702 &1&5020 &1&5018 &1&5016 
\\ 
{\tt WWGENPV}& 1&5700(4) &  \multicolumn{2}{c|}{}     &
\multicolumn{2}{c|}{}     &1&5014(4) 
\\ \hline 
\end{tabular}
\end{center}
\end{table}
\begin{table}[bhtp]
\caption[]{
\it
Results from different programs for the total {\tt CC11}
cross sections  $e^+e^- \rightarrow 4f$ and 
$e^+e^- \rightarrow 4f + \gamma $  (both in pbarn)
as functions of $\sqrt{s}$ for the $QQ' \, qq'$ channel.
\label{rfdp4}
}
\vspace{0.31cm}
\begin{center}
\begin{tabular}{|c|c|r@{.}l|r@{.}l|r@{.}l|}
\hline 
 & \hspace{.5cm}  $\sqrt{s}$ \hspace{.5cm}  
&\multicolumn{2}{c|}{\hspace{.5cm}   500 \hspace{.5cm} } 
&\multicolumn{2}{c|}{\hspace{.5cm}   1000 \hspace{.5cm} }
&\multicolumn{2}{c|}{\hspace{.5cm}   2000 \hspace{.5cm} }
\\ 
\hline 
\hline
$\sigma_{Born}$        & {\tt GENTLE}    
&  0&81482   & 0&32609    & 0&16684
\\                     & {\tt WWGENPV}   
&  0&81480(6)& 0&32602(6) & 0&16682(7) 
\\ \hline 
\hline
$\sigma_{QED}$ & {\tt GENTLE,SF} 
& 0&86950     &0&36514    &   0&18247
\\                     & {\tt WPHACT}
& 0&86956(9)  &0&36515(5) &   0&18250(4) 
\\                     & {\tt WTO}
& 0&86960(25) & \multicolumn{2}{c|}{} & \multicolumn{2}{c|}{} 
\\                     & {\tt WWGENPV}
& 0&86956(14) & 0&36530(35) & 0&18247(13) 
\\ \hline 
\end{tabular}
\end{center}
\end{table}
In table~2, a comparison of cross sections including universal ISR is
performed.
The high level of numerical and general precision is clearly
demonstrated.   

\section*{Acknowledgments}
We would like to thank the authors of the Fortran programs {\tt
  WPHACT}, {\tt WWGENPV}, {\tt WTO} for their kind cooperation when
the numerical comparisons have been performed.
%

 
\end{document}